\documentstyle[epsfig]{l-aa}

\begin{document}

  \thesaurus{3           
             (12.03.4    
              11.17.1)   
            }
   \title{The fine-structure constant:
         a new observational limit on its cosmological
         variation and some theoretical consequences}

   \author{A.V. Ivanchik, A.Y. Potekhin, D.A. Varshalovich
          }

   \offprints{A.V. Ivanchik, iav@astro.ioffe.rssi.ru}

   \institute{Department of Theoretical Astrophysics,
              Ioffe Physical-Technical Institute,
              St.-Petersburg 194021, Russia\\
             }

   \date{Received xxxxxxx 00, 1997; accepted xxxxxx 00, 0000}

   \maketitle

 \markboth{A.V.~Ivanchik et al.: The fine-structure constant}{
 A.V.~Ivanchik et al.: The fine-structure constant}

 \begin{abstract}

   Endeavours of the unification of the four fundamental
   interactions have resulted in a development of theories
   having cosmological solutions in which low-energy limits
   of fundamental physical constants vary with time. The validity of such
   theoretical models should be checked by comparison of the theoretical
   predictions with observational and experimental bounds on possible
   time-dependences of the fundamental constants.

   Based on high-resolution measurements of quasar spectra, we obtain
   the following direct limits on the average rate of the cosmological
   time variation of the fine-structure constant $\alpha$:
   $$ \left|\;\dot{\alpha}/\alpha\,\right|
                  <1.9\times10^{-14}{\rm~yr}^{-1} $$
   is the most likely limit, and
   $$ \left|\;\dot{\alpha}/\alpha\,\right|
                  <3.1\times10^{-14}{\rm~yr}^{-1} $$
   is the most conservative limit.

   Analogous estimates published previously, as well as other
   contemporary tests for possible variations of $\alpha$
   (those based on the ``Oklo phenomenon'', on the primordial
   nucleosynthesis models, and others) are discussed and compared
   with the present upper limit. We argue that the present result
   is the most conservative one.

 \end{abstract}

\section{Introduction}

   The fine-structure constant $\alpha=e^2/ \hbar c$ is the key
   parameter of Quantum Electrodynamics (QED). Initially it was
   introduced by Sommerfeld in 1916 for describing the fine structure 
   of atomic levels and corresponding resonance lines. Later it
   became clear that $\alpha$ is important for description of the 
   gross structure of atomic and molecular spectra as well as fine one.
   Moreover, now we know that {\it any electromagnetic 
   phenomena may be described in terms of the powers of $\alpha$}.
   In reality, $\alpha$ is not a true constant. It is established
   in the quantum field theory and confirmed by high-energy experiments
   that the coupling constants depend on distance (or momentum, or
   energy) because of vacuum polarization (see, e.g., Okun 1996). Here
   we consider the low-energy limit of $\alpha$, namely, its variation
   in the course of the cosmological evolution of the Universe.
   The possibility of fundamental constants to vary arose in the
   discussion of Milne (1937) and Dirac (1937). (For a sketch of the
   history of the problem of variability of the fundamental constants,
   see Varshalovich \& Potekhin 1995). The most important Dirac's 
   statement was that 
   {\it the constancy of the fundamental physical constants should be
   checked in an experiment}.

   The value of the fine-structure constant is known with rather
   high accuracy $\sim 4\times 10^{-9}$. The CODATA recommended value
   based on the 1997 adjustment of the fundamental constants of
   physics and chemistry is equal to $\alpha = 1/137.03599993(52)$.
   However, even this high precision does not exclude the possibility
   that the $\alpha$ value was different in early cosmological epochs.
   Moreover, some contemporary theories allow $\alpha$ to be different
   in different points of the space-time. Endeavours of unification of
   all fundamental interactions lead to a development of
   multidimensional theories like Kaluza-Klein and superstring ones
   which predict not only energy dependence of the
   constants but also dependence of their low-energy limits on
   cosmological time.

 {\it Superstring theories.} The superstring theory is a real
   candidate for the theory which is able to unify gravity with all
   other interactions. At present this theory is the only one which
   treats gravity in a way consistent with quantum mechanics. In the
   low energy limit $(E \ll E_{\rm Planck}\equiv\sqrt{\hbar c^5/G}
   \simeq 1.2\times10^{19}$~GeV, where $G$ is the gravitational constant,
   $\hbar$ is the Planck constant, and $c$ is the speed of light),
   the superstring theory
   reduces to the classical General Relativity (GR) with, however, an
   important difference: All versions of the theory predict the
   existence of the dilaton which is a scalar partner of the tensorial
   graviton. This immediately leads to an important conclusion that
   all the coupling constants and masses of elementary particles, being
   dependent on the dilaton scalar field $\phi$, should be space and
   time dependent. Thus, the existence of a weakly coupled massless
   dilaton entails small, but non-zero, observable consequences
   such as Jordan-Brans-Dicke-type deviations from GR and cosmological
   variations of the fine structure constant and other gauge
   coupling constants. The relative rate of the $\alpha$ variation
   is given by
   \begin{equation}
           \hspace{2.7cm}
           \frac{\dot \alpha}{\alpha} \sim kH(\phi-\phi_m)^2
   \label{Damour}
   \end{equation}
   where $k$ is the main parameter that determines the efficiency of
   the cosmological relaxation of the dilaton field $\phi$ towards
   its extreme value $\phi_m$, and $H$ is the Hubble constant 
   (Damour \& Polyakov 1994).
   In principle, the variation of $\alpha$ defined by Eq.~(\ref{Damour}) depends
   on cosmological evolution of the dilaton field and may be
   non-monotonous as well as different in different space-time regions.

 {\it Kaluza-Klein theories.}
   In superstring models the transition from $(4+D)$-dimensional string
   objects to the four-dimensional observed reality proceed through
   a compactification of the extra dimensions.
   Generalised Kaluza-Klein theories offer another possibility
   of unification of gravity and the other fundamental gauge
   interactions via their geometrization in a $(4+D)$-dimensional
   curved space. In such theories the truly fundamental constants are
   defined in $4+D$ dimensions and any cosmological evolution of the
   extra $D$ dimensions would result in a variation of the
   fundamental constants measured in the observed four-dimensional
   world. For example, in the Kaluza-Klein theories the fine-structure
   constant $\alpha$ evolves as
   \begin{equation}
    \hspace{3.47cm} \alpha \propto R^{-2}
   \label{R-2}
   \end{equation}
   where $R$ is a geometric scaling factor which characterises the
   curvature of the additional $D$-dimensional subspace (Chodos \& 
   Detweiler 1980; Freund 1982; Marciano 1984).

   There are many different versions of the theories described above
   and they predict various time-dependences of the fundamental constants.
   Thus, {\it bounds on the variation
   rates of the fundamental constants may serve as an important tool
   for checking the validity of different theoretical models of the
   Grand Unification and cosmological models related to them.}

   In the next section we briefly consider the basic methods allowing
   one to obtain restrictions on possible variations of fundamental
   constants. In Sect.~\ref{sect-astro}, an astrophysical method
   is considered in more detail. In Sect.~\ref{sect-crit}, we specify the
   properties of observational data required to reach a high accuracy
   of evaluation of $\dot{\alpha}/\alpha$.
   In Sect.~\ref{sect-obs}, we describe an observational technique
   employed to obtain data of the required quality.
   An upper bound on $|\dot{\alpha}/\alpha|$ is presented 
   and its theoretical consequences are discussed in Sect.~\ref{sect-res}.
   In Sect.~\ref{sect-concl}, conclusions are given.

\section{Characteristic features of methods for checking possible
         variations of the fundamental constants}

   Theoretical and experimental techniques used to investigate time
   variation of the fundamental constants may be divided into
   extragalactic and local methods. The latter ones include
   astronomical methods related to the Galaxy and the Solar system,
   geophysical methods, and laboratory measurements.

   An interest in the problem of changing constants has increased after
   an announcement  of a relative frequency drift observed by several
   independent research groups using long term comparisons of the
   Cs frequency standard with the frequencies of H- and Hg$^+$-masers.
   Such drift, in
   principle, could arise from a time variation of $\alpha$ because 
   H-maser, Cs, and Hg$^+$ clocks have a different dependence on $\alpha$
   via relativistic contributions of order $(Z\alpha)^2$. A detailed
   description of the laboratory method based on such comparisons may
   be found in Prestage et al.\ (1995), Breakiron (1993).

   Stringent limits to $\alpha$ variation have been presented by
   Shlyakhter (1976) and Damour \& Dyson (1996), who have analysed the
   isotope ratio $^{149}$Sm/$^{147}$Sm produced by the natural uranium
   fission reactor that operated about $2\times 10^9$ yrs
   ago in the ore body of the Oklo site in Gabon, West Africa. This
   ratio turned out to be considerably lower than that in the natural
   samarium, which is believed to have occurred due to the neutron
   capture by $^{149}$Sm during the uranium fission. Shlyakhter (1976)
   has concluded that the neutron capture cross section in
   $^{149}$Sm has not changed significantly in the $2\times 10^9$ yrs.
   However, the rate of the neutron capture reaction is sensitive to
   the position of the resonance level $E_r$, which depends on the
   strong and electromagnetic interactions.
   At variable $\alpha$ and invariable constant of the strong
   interaction (that is just a model assumption), the shift of
   the resonance level would be determined by changing the difference of
   the Coulomb energies between the ground state $^{149}$Sm and 
   the excited state of $^{150}$Sm$^*$ corresponding to the resonance
   level $E_r$. Unfortunately, there are no experimental data for
   the excited level of $^{150}$Sm$^*$ in question. 
   Damour \& Dyson (1996) have assumed that the Coulomb energy difference
   between the nuclear states in question is not less than one between
   the {\em ground\/} states of $^{149}$Sm and $^{150}$Sm. The latter
   energy difference has been estimated from isotope shifts and equals
   $\approx 1$ MeV. However, it looks unnatural that a weakly bound
   neutron, captured by a $^{149}$Sm nucleus to form the highly excited
   state $^{150}$Sm$^*$, can so strongly
   affect the Coulomb energy. Moreover, the data of isomer shift
   measurements of different nuclei indicate that the mean-square radii
   (and therefore the Coulomb energy) of the charge-distribution for
   excited states of heavy nuclei may be not only larger but also
   considerably {\em smaller\/} than the corresponding radius for the
   ground states (Kalvius \& Shenoy 1974). This indicates the
   possibility of violation of the basic assumption involved by Damour
   \& Dyson (1996) in the analysis of the Oklo phenomenon, and therefore
   this method may possess a lower actual sensibility.

   Another possibility of studying the effect of changing fundamental
   constants is to use the standard model of the primordial
   nucleosynthesis. The amount of $^4$He produced in the Big Bang is
   mainly determined by the neutron-to-proton number ratio at the
   freezing-out of n$\leftrightarrow$p reactions. The freezing-out
   temperature $T_f$ is determined by the competition between the
   expansion rate of the Universe and the $\beta$-decay rate. A
   comparison of the observed primordial helium mass fraction,
   $Y_p=0.24\pm0.01$, with a theoretical value allows to obtain
   restrictions on the difference between the neutron and proton
   masses at the epoch of the nucleosynthesis and, through it,
   to estimate relative variation of the curvature radius $R$ of the
   extra dimensions in multidimensional Kaluza--Klein-like theories
   as well as the $\alpha$ value (Kolb et al. 1986; Barrow 1987).
   However one can notice that different coupling constants might
   change simultaneously. For example, increasing 
   the constant of the weak interactions $G_F$ would cause
   a weak freeze-out at a lower temperature, hence a decrease in
   the primordial $^4$He abundance. This process would compete with
   the one described above, therefore, it reduces sensibility of the
   estimates. Finally, the restrictions would be
   different for different cosmological models since the expansion
   rate of the Universe depends on the cosmological constant $\Lambda$.

   A number of other methods are based on the stellar and planetary
   models. The radii of the planets and stars and the reaction rates
   in them are influenced by values of the fundamental constants,
   that offers a possibility to check the variability of the constants
   by studying, for example, lunar and Earth's secular accelerations,
   which has been done using satellite data, tidal records, and ancient
   eclipses. Another possibility is offered by analysing the data
   on binary pulsars and the luminosity of faint stars. A variety of
   such methods has been critically reviewed by Sisterna \& Vucetich
   (1990). All of them have relatively low sensibility.

   An analysis of natural long-lived
   $\alpha$- and $\beta$- decayers, such as $^{187}$Re or $^{149}$Sm
   in geological minerals and meteorites,
   is much more sensitive (e.g., Dyson 1972).
   A combination of these methods, reconsidered by Sisterna
   \& Vucetich (1990), have yielded an estimate
   $\dot{\alpha}/\alpha = (-1.3 \pm 6.5)\times 10^{-16}\ {\rm yr}^{-1}$.

   The following weak points are inherent to the methods described above:
   \begin{itemize}
   \item[(i)] The derived restrictions strongly depend on model
              conjectures.
   \item[(ii)] The ``local'' methods give estimates for only
         a narrow space-time region around the Solar system.
         For example, the time of the operation of the Oklo reactor
         corresponds to the cosmological redshift $z \approx 0.1$,
         while a method based on observations of quasar spectra 
         (described below) makes
         it possible to bring out evolutionary effects for large span
         of redshifts to $z \sim 5$.
   \end{itemize}

   This and other problems, as we believe, can be solved by an
   astrophysical method which will be described in the next section.

   \begin{figure}[t]
       \begin{center}
       \epsfysize=10.7cm
       \epsfbox{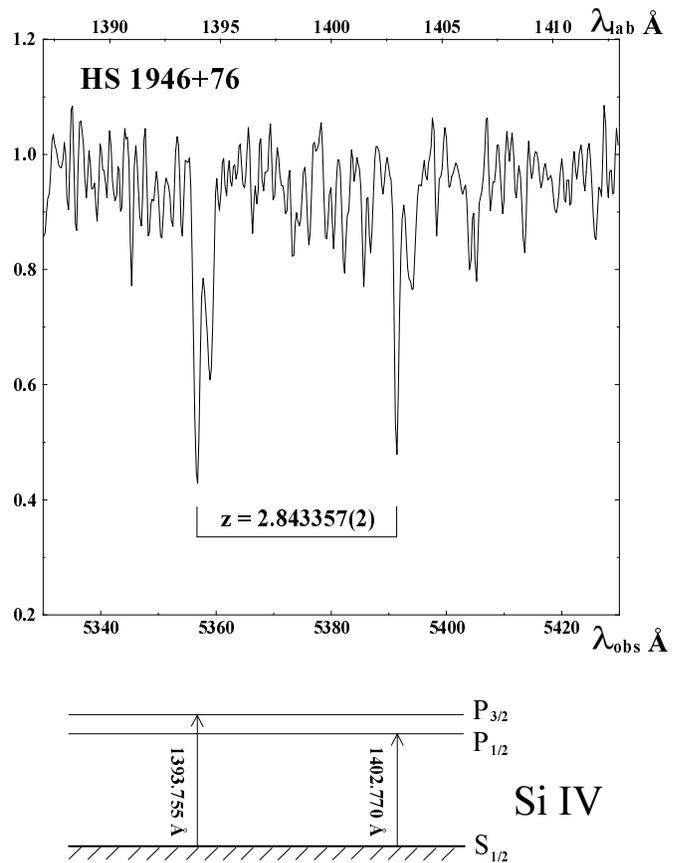}
       \end{center}
      \caption{Fine Doublet Splitting from Quasar Spectra}
         \label{FigSi}
   \end{figure}

\section{The astrophysical method\label{sect-astro}}
   An astrophysical method that allows one to estimate $\alpha$ value
   at early stages of the Universe evolution was originally proposed
   by Savedoff (1956). Bahcall and Schmidt (1967) were the first to apply
   this method to fine-splitting doublets in quasar spectra. A recent
   update of this method has given the result:
   $(-4.6 \rightarrow 4.2) \times 10^{-14} \;$ yr$^{-1}$ for upper
   limit (95\% C.L.) on a possible $\alpha$ variation
   (Cowie and Songaila 1995). A more stringent limit is presented in
   the paper by Varshalovich et al.\ (1996) and in this paper.

   We adhere the conventional belief that the quasar is an object of
   cosmological origin.
   Its continual spectrum was formed at an epoch corresponding to the
   redshift $z$ of main emission details specified by relationship
   $\lambda_{\rm obs} = \lambda_{\rm lab}(1+z)$.

   Quasar spectra show the absorption resonance lines of the ions
   C IV, Mg II, Si IV, and others, corresponding to the
   $S_{1/2} \rightarrow P_{3/2} \; (\lambda_1)$ and
   $S_{1/2} \rightarrow P_{1/2} \; (\lambda_2)$ transitions
   (see Fig.~\ref{FigSi}). 
   The difference between $\lambda_1$ and $\lambda_2$ is
   due to the fine splitting between the $P_{3/2}$ and $P_{1/2}$ energy
   levels. The relative magnitude of the fine splitting of the
   corresponding resonance lines is approximately proportional to the
   square of $\alpha$,
   \begin{equation}
    \hspace{3.6cm} \frac{\delta \lambda}{\lambda} \propto \alpha^2,
   \label{delta-lambda}
   \end{equation}
   where $\delta \lambda = \lambda_2 - \lambda_1$ and
   $\lambda = (\lambda_2+\lambda_1)/2$. Thus, the ratio
   $(\delta \lambda / \lambda)_z / (\delta \lambda / \lambda)_0$
   is equal to $\left(\alpha_z / \alpha_0\right)^2$, where the
   subscripts $z$ and 0 denote the fine-splitting doublet in a quasar
   spectrum and the laboratory value, respectively. Hence, the relative
   change in $\alpha$ can be approximately written as
   \begin{equation}
    \hspace{2.2cm}
    \left(\frac{\Delta \alpha}{\alpha} \right)_z = \frac{1}{2}
    \left[\frac{( \delta \lambda / \lambda)_z}
               {( \delta \lambda / \lambda)_0} - 1 \right],
   \label{delta-alpha}
   \end{equation}
   provided that $(\Delta \alpha / \alpha)$ is small.

   Thus, by measuring $\lambda_1$ and $\lambda_2$ in an absorption system
   corresponding to the redshift $z$ and comparing the derived and
   laboratory values, one can directly estimate the difference between
   $\alpha$ at the epoch $z$ and the present value.

   It is to be noted that this method is the only one which provides
   direct values of $\Delta \alpha / \alpha$ for different space-time
   points of the Universe. Thus, using this method we can study not
   only possible deviations of the fine-structure constant from its 
   present value but also its values in space regions which were 
   causally disconnected at earlier evolutionary stages of the 
   Universe (e.g. Tubbs \& Wolfe 1980; Varshalovich \& Potekhin 1995).

   Tubbs \& Wolfe (1980) (see also references therein) have used 
   a coincidence of redshifts of optical resonant lines of ions with 
   redshifts of the hydrogen 21 cm radio lines in distant absorption 
   systems to derive an upper limit on the combination 
   $\alpha^2 g_{\rm p} m_{\rm e}/m_{\rm p}$ at $z\sim 0.4-1.8$.
   Here, $g_p$, $m_{\rm e}$ and $m_{\rm p}$ are the proton gyromagnetic 
   factor and the masses of electron and proton, respectively.
   Recently, Drinkwater et al. (1998) used a similar method of comparison
   of the redshifts of the HI (21 cm) and molecular radio lines
   at redshift $z=0.68$ in two quasars to place a new strongest 
   restriction on fractional variation of $g_p \alpha^2$ at the level 
   $\sim 10^{-15}{\rm~yr}^{-1}$, from which they concluded that
   $|\dot\alpha/\alpha|<5\times10^{-16}{\rm~yr}^{-1}$.
   Two drawbacks are inherent to this method of comparison:
   (i) It does not give direct restriction on variation of $\alpha$,
   $g_p$, or electron-to-nucleon mass ratios, but only on
   their combination.
   (ii) One cannot be sure that the hydrogen and molecular absorption 
   lines originate from the same cloud along the respective lines of sight.
   Since the analysed line profiles were complex and required 
   decomposition into several contours, the perfect juxtaposition of 
   the hydrogen and molecular lines (hence the seemingly extra high 
   accuracy) might result from accidental coincidence of redshifts of 
   different components of the line profiles. Only analysis of the ratios
   of wavelengths of the same ion species (in particular, the doublet 
   ratios mentioned above) is free of such ambiguity.

Detailed discussion of possible sources of systematical 
and statystical errors 
of the method based on Eq.~(\ref{delta-alpha})
has been given elsewhere (Potekhin \& Varshalovich 1994).
The most significant source of possible systematical error
turned out to be the uncertainty in the laboratory wavelengths $\lambda_0$
(see below). Some errors which have a systematic character
for one selected absorption system
(e.g., the shifts of estimated line centers
resulting from occasional unidentified blending
of partially saturated lines) become random (statistical)
for a sample of unrelated absorption systems.
For this reason, Potekhin \& Varshalovich have argued
that one should not confide in the errorbars
estimated for individual absorption systems,
but derive the error
from the actual scatter of the data.
For example, Webb et al. (1998), using individual 
estimates of statistical errors of several fine-structure
absorption systems, reported an unprecedented 
accuracy and claimed an evidence for nonmonotonic $\alpha$ variation:
$\Delta\alpha/\alpha=(-2.64\pm0.35)\times10^{-5}$ at $1.0<z<1.6$.
However, an independent statistical treatment of the data
presented in their paper, disregarding the errorbars but using the
actual scatter of the data to estimate the confidence level, 
yields (in the same range of redshift)
$\Delta\alpha/\alpha=(-3.0\pm1.0)\times 10^{-5}\pm\sigma_{\rm syst}$,
indicating that the individual statistical errors 
reported by the authors significantly underestimate
the actual statistical uncertainty.
Here, the systematic error $\sigma_{\rm syst}\sim10^{-4}$ 
is mainly due to the uncertainty
in the laboratory wavelengths [see Eq.~(\ref{sigma0}) below].

In this paper we apply the described method to a sample
of 20 absorption systems at $2.0<z<3.6$,
selected according to the criteria formulated in the next section.

\section{Criteria of data selection\label{sect-crit}}

   Of the available alkali-like doublets observed in quasar spectra
   such as C IV, Mg II, Al III, O VI, N V, Si IV, and others,
   we have selected for our analysis the Si IV line doublet, because
   it has the greatest ratio
   $\delta \lambda / \lambda = 6.45 \times 10^{-3}$, allowing this
   ratio to be measured most accurately. 
   The abundance of silicon and its
   ionization state, as a rule, is such that the Si IV doublet lines
   occur on a linear part of the growth curve, which simplifies
   determination of the central wavelength of each line.

   The laboratory values of the Si IV doublet
   wavelengths ($\lambda_{01}=1393.755$~\AA\ and 
   $\lambda_{02}=1402.769$~\AA, according to Striganov \& Odintsova 1982;
   $\lambda_{01}=1393.755$~\AA\ and 
   $\lambda_{02}=1402.770$~\AA, according to Morton et al.\ 1988)
   are known with an uncertainty $\sigma_{\lambda_0}\approx 1$~m\AA.
   This uncertainty can introduce an appreciable systematic error in
   the determination of $(\Delta \alpha / \alpha)_z$
   from Eq.~(\ref{delta-alpha}).
   In the case of the Si IV doublet, this error can be estimated as
   \begin{equation}
      \hspace{1.4cm}
      \sigma_{\rm syst}(\Delta\alpha/\alpha)\approx
      \sigma_{\lambda_0}/(\delta \lambda \sqrt{2})\approx 8\times 10^{-5}.
   \label{sigma0}
   \end{equation}
   This value of the systematic error for Si IV is the smallest
   among all alkali-like ion species listed above.

   Available observational data on Si IV doublets in quasar spectra
   have been selected taking into account the following criteria:
   \begin{enumerate}
   \item  High resolution FWHM $<$ 25 km/s.
   \item  Wavelength of each doublet component is measured
          independently (without including the a priori information
          on $\delta\lambda/\lambda$).
   \item  Any obvious blending of each component of the doublet
          is absent.
   \item  Equivalent width ($W_{1,2}$) ratio of the doublet
          components is limited to $\;1 \leq (W_1 \pm 2 \sigma_{W_1})
          /(W_2 \pm 2 \sigma_{W_2}) \leq 2$ \\
          (in conformity with the oscillator strength ratio).
   \item  The lines are detected at the level higher than
          $5 \sigma$, where $\sigma$ is a noise level.
   \item  A dispersion curve for calibration is carefully determined
          -- e.g., using the technique described in the next section.
   \end{enumerate}

   Unfortunately, many of recent data found in literature do not
   often satisfy the necessary conditions 2, 3, and 5.

\section{Observations and data reduction\label{sect-obs}}

   Suitable observational data have been presented by Petitjean
   et al.\ (1994), Cowie \& Songaila (1995), and Varshalovich
   et al.\ (1996). To give an idea about basic elements of
   observations and data reduction we describe here the main stage
   of the work by Varshalovich et al.\ (1996).

   Observations were carried out in 1994 using the Main Stellar
   Spectrograph of the 6-m Telescope. A Schmidt camera (F:2.3),
   reconstructed to operate with a CCD array, was employed. In
   combination with a 600 lines mm$^{-1}$ diffraction grating
   operating in the second order, this camera yielded a linear
   reciprocal dispersion of 14 \AA $\;$ mm$^{-1}$
   (0.24 \AA\ pixel$^{-1}$). Since this study is metrological
   in essence, special attention was paid to wavelength calibration
   and check measurements. A Th+Ar lamp and the Atlas of the
   Th+Ar spectrum constructed from echelle data (D'Odorico et al.
   1987), which gave vacuum wavelengths $\lambda$ to within
   0.001 \AA, were employed for calibration. Ten to twenty reference
   lines were used to construct the dispersion curve in the form of
   a Chebyshev polynomial ($\lambda=a_0T_0+a_1T_1+a_2T_2+...$,
   where $T_n$ is the Chebyshev polynomial of $n$th order) for each
   of the spectral segments in question. 
   The nonlinear terms ($a_2$) of the dispersion curve
   were determined especially carefully, because
   the accuracy of the measured ratio $\delta \lambda / \lambda$ was
   particularly sensitive to the nonlinearity of the scale, contrary
   to the absolute calibration (i.e. $a_0$). For the
   same reason, the spectrograph was adjusted for each of the Si IV
   doublets in such way that both doublet components, with a separation
   of about 36 \AA\ for $z \sim 3$, were in the middle of the
   spectral segment corresponding to one diffraction order, where the
   nonlinear distortions of the scale were at a minimum, while the
   sensitivity was at a maximum.

\begin{figure}[t]
      \begin{center}
      \epsfysize=5.9cm
      \epsfbox{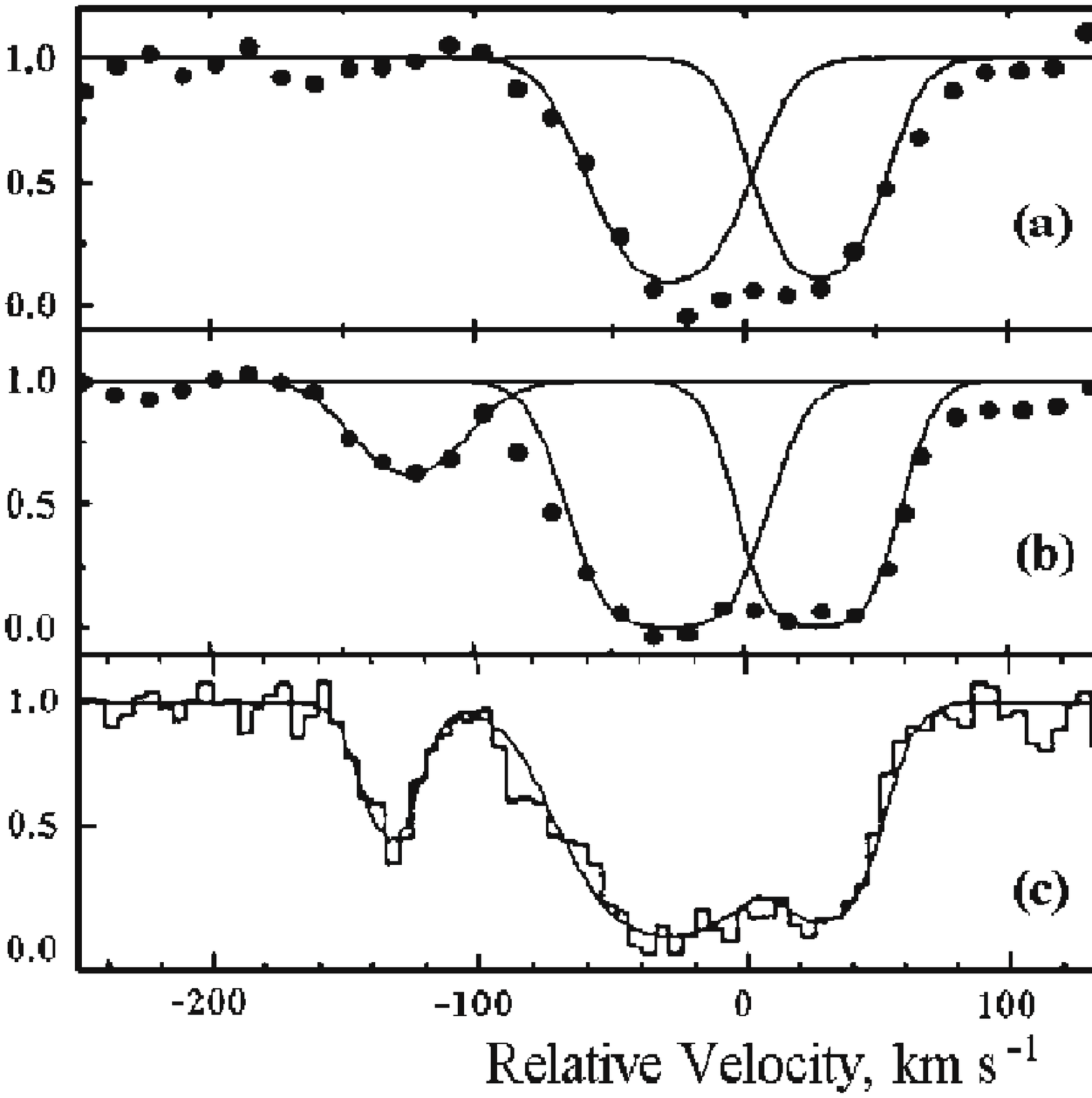}
      \end{center}
      \caption{The spectrum of HS 1946+76: comparison of the profiles
      of the Si IV doublet absorption lines with $z_{\rm abs} = 3.049$
      [the $\lambda_0 = 1393$ \AA $\;$ (a) and 1402 \AA $\;$ (b) lines
      in our study and the $\lambda_0 = 1402$ \AA $\;$ line (c)
      in Fan and Tytler (1994)]
   \label{profile}}
   \end{figure}

The wavelengths $\lambda_1$ and $\lambda_2$
were determined by gausian decomposition of the corresponding 
absorption lines in the observed quasar spectra.
   Portions of the spectrum of HS 1946+76 ($z_{\rm em} = 3.05$), one
   of the brightest QSOs in the sky, measured by Varshalovich et
   al.\ (1996), are shown in Figs.~\ref{FigSi} and \ref{profile}.
Despite the absorption doublet shown in Fig.~\ref{FigSi} consists
of two components, it still satisfies the criteria formulated in 
Sect.~\ref{sect-crit}, including item 3.
In fact, we have checked that even total neglect of the second
component of each line alters the estimate
of $\Delta\alpha/\alpha$ for this absorption system
by less than $2\times10^{-5}$, which is well below
the overall statistical uncertainty reported in Eq.~(\ref{estimate}) below.
On the other hand, the portion of spectrum shown in Fig.~\ref{profile}
requires caution, since the spectral components are close
and partially saturated.

   In their recent analysis of the spectrum of HS 1946+76, Fan and
   Tytler (1994) have detected six absorption systems, two of which
   contain Si IV lines. However, they presented only one of the doublet
   components in each of the doublets. Therefore, their high-resolution
   results cannot be used for measuring $\Delta \alpha$.
   In Fig.~\ref{profile}, the profiles of our
   measured spectral lines are compared with the profile obtained by 
   Fan \& Tytler with resolution of $\sim 10$ km s$^{-1}$.
The comparison confirms that the resolution achieved in our work
is equally sufficient for the profile analysis.

\section{Results\label{sect-res}}

\subsection{Bounds on $\alpha$\label{sect-bounds}}

   Results of the analysis of the Si IV fine-splitting doublet
   lines are presented in Table 1. The third column shows a deviation
   of the $\alpha$ value calculated according to Eq.~(\ref{delta-alpha}) 
   for a single doublet. The value 
   $(\delta\lambda/\lambda)_0=0.0064473$ in Eq.~(\ref{delta-alpha})
   has been adopted from Morton et al.\ (1988). From the data listed in 
   Table~1, $(\Delta \alpha/\alpha)$ is estimated by standard least squares.
According to the arguments given in Sect.~\ref{sect-astro},
we did not involve estimates of the individual wavelength
uncertainties. For example, for the profile shown in Fig.~\ref{FigSi},
the uncertainty of the gaussian decomposition yields $\sigma_\lambda\sim
0.007$\,\AA, and the estimate $\sigma_\lambda$ as a function of resolution
and signal-to-noise ratio, following Young et al. (1979),
yields even smaller $\sigma_\lambda\sim
0.005$\,\AA, while the actual uncertainty may be larger. 
   Estimating the 
statistical error from the scatter of $\Delta\alpha/\alpha$ measurements
and taking into account the systematic error 
   in Eq.~(\ref{sigma0}), we arrive at the final estimate (for $z$ interval
   $2.0-3.5$):
   \begin{equation}
     \hspace{0.7cm} \left(\Delta \alpha / \alpha \right) =
                         (-3.3 \pm 6.5 {\rm~[stat]}
                               \pm 8.0 {\rm~[syst]}) \times 10^{-5}.
    \label{estimate}
   \end{equation}
   The corresponding bound at the 95\% significance level is
   \begin{equation}
      \hspace{2.6cm}
      \left|\Delta \alpha/\alpha \right|
    < \varepsilon = 2.3 \times 10^{-4}.
    \label{fin-res}
   \end{equation}

   Since we have considered a sample of 20 absorption systems, 
the statistical (observational) error in Eq.~(\ref{estimate}) 
is $\sim \sqrt{20}\approx4.5$ times smaller than a typical 
error of a single measurement.
On the other hand, the statistical error and
   the systematic error due to the above-mentioned uncertainty of the
   laboratory wavelengths are of the same order of magnitude, hence any
   significant further improvement of the upper limit (\ref{fin-res}) 
   would require an improvement of not only observational techniques
   but also the accuracy of the laboratory measurements.

   According to the standard cosmological model with the 
   cosmological constant $\Lambda=0$, the time elapsed since
   the formation of the absorption spectrum with redshift $z$ is
   \begin{eqnarray}
    \hspace{1.7cm}
   t_z&=&t_0[1-(1+z)^{-3/2}] \quad  (\Omega=1), \\
   t_z&=&t_0[1-(1+z)^{-1}] \quad  \;\; (\Omega \ll 1),
   \end{eqnarray}
   where $\Omega$ is the mean-to-critical density ratio.
   With the present age of the Universe equal to
   $(14 \pm 3) \times 10^9$ yr, we see that the redshifts
   $\,z=2.8 \pm 0.7 \;$ presented in Table 1 correspond to
   the elapsed time $\approx 12 \pm 4$ Gyr.

   Thus, the upper limit on the time-averaged
   rate of change of $\alpha$ (with most likely values
   $z=2.9$ and $t_0=14\times 10^9$ yr) at the $2\sigma$ level is
   \begin{equation}
     \hspace{2.4cm}
     \left| \overline{\dot{\alpha}/\alpha} \right|
    < 1.9 \times 10^{-14} {\rm~yr}^{-1} .
   \end{equation}
   And the most conservative upper limit
   ($z=2.1$ and $t_0=11\times 10^9$ yr) at the $2\sigma$ level is
   \begin{equation}
     \hspace{2.4cm}
     \left| \overline{\dot{\alpha}/\alpha} \right|
    < 3.1 \times 10^{-14} {\rm~yr}^{-1} .
   \label{alpha-dot}
   \end{equation}

   \begin{table}[t]
    \caption{ $\;\,$ Variation of $\alpha$ values estimated from redshifted $\;\;\;\;$
              Si IV doublet line splitting
              according to Eq.~(\protect\ref{delta-alpha})
              \label{tab1}}
     \[
      \begin{array}{r@{\hspace{9mm}}l@{\hspace{4mm}}r@{\hspace{5mm}}c}
         \hline
         \hline
         \noalign{\smallskip}
      {\rm Quasar}~~~ & ~~~~~z & (\Delta \alpha / \alpha)_z,10^{-4} & 
{\rm Ref.}  \\
         \noalign{\smallskip}
         \hline
         \noalign{\smallskip}
         $HS 1946+76$ & 3.050079 &  1.58 \;\;\;\;\;\;\; & 1  \\
         $HS 1946+76$ & 3.049312 &  0.34 \;\;\;\;\;\;\; & 1  \\
         $HS 1946+76$ & 2.843357 &  0.59 \;\;\;\;\;\;\; & 1  \\
         $S4 0636+68$ & 2.904528 &  1.37 \;\;\;\;\;\;\; & 1  \\
         $S5 0014+81$ & 2.801356 & -1.80 \;\;\;\;\;\;\; & 1  \\
         $S5 0014+81$ & 2.800840 & -1.70 \;\;\;\;\;\;\; & 1  \\
         $S5 0014+81$ & 2.800030 &  1.11 \;\;\;\;\;\;\; & 1  \\
                      &          &       \;\;\;\;\;\;\; &    \\
        $PKS 0424-13$ & 2.100027 & -4.51 \;\;\;\;\;\;\; & 2  \\
        $Q   0450-13$ & 2.230199 & -1.48 \;\;\;\;\;\;\; & 2  \\
        $Q   0450-13$ & 2.104986 &  0.02 \;\;\;\;\;\;\; & 2  \\
        $Q   0450-13$ & 2.066646 &  1.03 \;\;\;\;\;\;\; & 2  \\
                      &          &       \;\;\;\;\;\;\; &    \\
        $Q   0302-00$ & 2.785    &  2.07 \;\;\;\;\;\;\; & 3  \\
        $PKS 0528-25$ & 2.813    &  1.29 \;\;\;\;\;\;\; & 3  \\
        $PKS 0528-25$ & 2.810    &  1.03 \;\;\;\;\;\;\; & 3  \\
        $PKS 0528-25$ & 2.672    & -5.43 \;\;\;\;\;\;\; & 3  \\
        $Q   1206+12$ & 3.021    & -1.29 \;\;\;\;\;\;\; & 3  \\
        $PKS 2000-33$ & 3.551    & -3.88 \;\;\;\;\;\;\; & 3  \\
        $PKS 2000-33$ & 3.548    &  2.85 \;\;\;\;\;\;\; & 3  \\
        $PKS 2000-33$ & 3.332    &  5.95 \;\;\;\;\;\;\; & 3  \\
        $PKS 2000-33$ & 3.191    & -5.69 \;\;\;\;\;\;\; & 3  \\
         \noalign{\smallskip}
         \hline
      \end{array}
     \] 
  References:
    [1] Varshalovich et al.\ (1996);
    [2] Petitjean et al.\ (1994);
    [3] Cowie and Songaila (1995).
   \end{table}

\subsection{Consequences for theoretical models}

  The results obtained above make it possible to select a number of
  theoretical models for dependence of $\alpha$ on cosmological time
  $t$ (the time elapsed since the Universe creation).
  Consider some of them: \\
  a. The hypothesis of the logarithmic dependence
  \begin{equation}
   \hspace{2.4cm}
   \alpha^{-1} \approx \ln \left( \frac{t}{8 \pi 
   \tau_{\rm Planck}} \right).
  \end{equation}
  Here, $\tau_{\rm Planck} 
  \equiv \sqrt{\hbar G / c^5}=5.4\times10^{-44}$ sec is the Planck 
  time. Teller (1948) suggested this dependence on the basis of the 
  striking concurrence of the value $\alpha^{-1}=137.036$ 
  and value $\ln ( t_0 / 8 \pi \tau_{\rm Planck} ) \approx 137$ (note 
  that the form of the dependence and $t_0$ value is presented in the 
  modern interpretation).
  Later Dyson (1972) showed that it could be
  supported by considerations that follow from the method of
  renormalization in the QED. In the standard cosmological model with
  $\Lambda = 0$ and $\Omega=1$, this hypothesis gives rise to the
  following dependence
   \begin{equation}
    \hspace{2.4cm}
    \alpha_z=\frac{\alpha_0}{1-\frac32\alpha_0\ln(1+z)}
   \end{equation}
  For $\,z=2.8$, this yields $\alpha_z=1.015\,\alpha_0$, in obvious
  contradiction with inequality (\ref{fin-res}). 
  Thus, the logarithmic dependence
  can be excluded. Note that a more general dependence
  $\alpha^{-1}=C \ln(t/ \tau)$ with arbitrary $C$ and $\tau \ge 
  \tau_{\rm Planck}$
  has been ruled out already by Varshalovich \& Potekhin (1995). \\
  b. The power-law dependence (for standard model)
  \begin{equation}
  \hspace{3.7cm}
  \alpha \propto t^n 
  \end{equation}
  in the standard flat Universe this leads to the dependence
  $\alpha = \alpha_0 (1+z)^{-3n/2}$, from which, using
  Eq.~(\ref{fin-res}), we obtain the bound
  \begin{equation}
    \hspace{1.8cm}  
    |n|< 2 \varepsilon /[3 \ln(1+z)] = 1.1 \times 10^{-4}
  \end{equation}
  c. The Kaluza-Klein-like models often assume that compactification
  was occurring at the very early stages of Universe evolution and
  now the additional $D$ dimensions of the subspace have
  radius of curvature $R \sim l_{\rm Planck}
  \equiv \sqrt{\hbar G / c^3} = 1.6\times 10^{-33} \,$cm.
  Thus, determining of variations at such level is impossible at 
  present. On the other hand, there are theories in which the process 
  of the compactification can develop even now. The information about
  $R$ for such theories may be obtained from measurements of the 
  $\alpha$ variation. According to Eq.~(\ref{R-2}), the restriction on 
  the changing of the subspace scaling factor is
  \begin{equation}
  \hspace{2.7cm}
  \frac{\Delta R}{R_0} \approx \frac12 \frac{\Delta\alpha}{\alpha}
  \la 10^{-4}
  \end{equation}
  d. The recent version of the dilaton evolution proposed by
     Damour \& Polyakov (1994) in the frame of the string model
     has provided an expression connecting the variation of the
     fundamental constants:
  \begin{equation}
  \hspace{2.6cm}
  \frac{\dot{G}}{G} = \Theta k \frac{\dot{\alpha}}{\alpha}
                    \simeq 3332.5 k \frac{\dot{\alpha}}{\alpha} .
  \end{equation}
  Here, $k$ is a parameter of the theory, which fixes a type and speed
  of the evolution of the scalar field (i.e. dilaton field),
  $\Theta=2(\ln(\Lambda_s/mc^2))^2$ is a numerical coefficient,
  $\Lambda_s \simeq 5 \times 10^{17}$ GeV is a string mass scale, 
  and $m=1.661 \times 10^{-24}$~g is the atomic mass unit 
  ($mc^2=931.5$ MeV).
  The parameter $k$ regulates the character of the time variation:
  $G$ and $\alpha$ oscillate for $k \geq 1$, whereas for $k \ll 1$,
  these constants change monotonously. For the models with $k \geq 1$
  discussed by Damour \& Polyakov (1994), a measured bound on $\dot{G}/{G}$
  enables one to obtain a bound on $\dot{\alpha}/\alpha$. For models
  with $k \ll 1$, we can obtain the bound on $\dot{G}/{G}$ from 
  Eq.~(\ref{alpha-dot}):
  $\left| \dot{G}/{G}\right| \ll 3332.5\left| \dot{\alpha}/\alpha \right| 
  <1\times10^{-10}$~yr$^{-1}$.

\section{Conclusions\label{sect-concl}}

   We have formulated criteria for selection of high-quality doublet
   lines in quasar spectra, most suitable for the analysis of the
   large-scale space-time variation of the fine-structure constant.
   We have also described the calibration technique capable to provide
   the necessary metrological quality.

   Using the formulated criteria and described technique, we have
   performed the Si IV doublet analysis of quasar spectra, that has
   enabled us to set an upper bound on the rate of the possible
   cosmological time variation of the fine-structure constant.
   This limit is much more model-independent and hence more robust
   than those provided by the local tests, including the analysis of
   the Oklo phenomenon.

   One may anticipate that with development of observational spectroscopy
   the portion of material which satisfies the formulated criteria
   will rapidly increase in the near future. However, as shows the
   discussion in Sect.~\ref{sect-bounds}, this will not allow one
   to further tighten the bound on $\dot \alpha / \alpha$ 
   until an improvement of the laboratory wavelength accuracy.
   On the other hand, in frames of the hypothesis
   that $\alpha$ does {\em not\/} change at the present stage
   of the cosmological evolution, the progress
   of the observational spectroscopy offers an exciting opportunity
   to improve the accuracy of determination of the fine
   splitting $\delta\lambda/\lambda$ against that
   available in the laboratory experiments.

\begin{acknowledgements}
We are grateful to J.D.~Barrow, 
C.W.~Churchill, 
V.A.~Dzuba, M.J.~Drinkwater, 
V.V.~Flambaum, and J.K.~Webb for providing us 
with preprints of their papers prior to publication.
We thank the referees, P.~van Baal and B.~H. Foing, for useful
suggestions which helped us to improve the manuscript.
   This work was supported by the Russian State Program
   ``Fundamental Metrology'' and Russian Foundation for Basic Research
   (grant 96-02-16849a).
\end{acknowledgements}

\end{document}